Non-alignment comparison of human and high primate genomes

*Kirzhner V.M., Frenkel S., Korol A.B.

Institute of Evolution and Department of Evolutionary & Environmental Biology,

University of Haifa, Mount Carmel, Haifa 31905, Israel

*Corresponding author: Kirzhner V.M., +972-48288040, valery@research.haifa.ac.il

# Abstract

Compositional spectra (CS) analysis based on k-mer scoring of DNA sequences was employed in this study for dot-plot comparison of human and primate genomes. The detection of extended conserved synteny regions was based on continuous fuzzy similarity rather than on chains of discrete anchors (genes or highly conserved noncoding elements). In addition to the high correspondence found in the comparisons of whole-genome sequences, a good similarity was also found after masking gene sequences, indicating that CS analysis manages to reveal phylogenetic signal in the organization of noncoding part of the genome sequences, including repetitive DNA and the genome "dark matter". Obviously, the possibility to reveal parallel ordering depends on the signal of common ancestor sequence organization varying locally along the corresponding segments of the compared genomes. We explored two sources contributing to this signal: sequence composition (GC content) and sequence organization (abundances of k-mers in the usual A,T,G,C or purine-pyrimidine alphabets). Whole-genome comparisons based on GC distribution along the analyzed sequences indeed gives reasonable results, but combining it with k-mer abundances dramatically improves the ordering quality, indicating that compositional and organizational heterogeneity comprise complementary sources of information on evolutionary conserved similarity of genome sequences.

**Key words:** k-mer abundance; fuzzy similarity; dot-plot; conserved synteny, noncoding DNA; genome coverage; phylogenetic signal; compositional vs. organizational heterogeneity

# Introduction

The fast-growing set of sequenced genomes is a unique resource for comparative studies of organization and evolution of genetic material via whole-genome sequence comparisons complementing the classical gene-centric approach (Karlin and Ladunga 1994; Kirzhner et al. 2002; Höhl and Ragan 2007; Kirzhner et al. 2007; Mrázek 2009). The objectives of such comparisons may include: (a) determination of evolutionary steps resulting in changes of compared genomes relative to the ancestor genome and reconstruction of the ancestor genome organization (Alekseyev and Pevzner 2009); (b) searching of the karyotype evolutionary breakpoints (Kemkemer et al. 2009); (c) searching for conserved genomic regions as a method for discovery of functional (e.g., regulatory) elements (Kellis et al. 2004; Taylor et al. 2006); (d) comparing different assemblies of the same genome resulting from different sequencing data or different assembly algorithms (Meader et al. 2010). A widespread approach for genome-wide comparisons is based on anchoring, in which orthologous genes or evolutionary conserved noncoding regions can be employed as anchors (Schwartz et al. 2003; Mahmood et al. 2010). The analysis includes detection of anchor regions in the compared genomes followed by the chaining of selected collinear non-overlapping anchors and, if possible (e.g., for closely related species), aligning the regions between the anchors (Abouelhoda and Ohlebusch 2005; Hachiya et al. 2009).

Numerous algorithms, alignment or alignment-free, were proposed for the fast detection of anchors between the compared genomes. Alignment-free methods of mapping orthologous regions of the compared genomes, e.g. by using k-mer (or "seed") enrichments as anchors of conserved synteny blocks, are especially attractive allowing to deal with both coding and noncoding genome sequences in a unified way (Schwartz et al. 2003; Kiełbasa et al. 2011). It is noteworthy, that revealing a highly conserved chain of anchors does not necessarily guarantee conservation of sequence stretches between the anchors. Here we present a new variant of an alignment-free genome comparison based on our recent modification of the dot-plot scheme (Kirzhner et al. 2011). It utilizes compositional spectra (CS) analysis (Kirzhner et al. 2002; Bolshoy et al. 2010) that is also based on k-mer scoring in the compared DNA sequences. Unlike the standard approaches based on seed-and-extend anchoring, our strategy employs the distance matrix that includes pair-wise distances of each-versus-each segments of the compared genomes.

Therefore, the decisions on conserved synteny are based on distributed (fuzzy) similarity rather than on chains of discrete (usually rather short) anchors. The proposed algorithm is illustrated by genome-wide analysis of primate sequences.

We divided the entire genome into segments of million bp each, and evaluated the similarity of these segments based on such simple characteristics as their GC content and composition spectra. We targeted four types of sequences: (a) the entire genome sequence; (b) the sequence with masked repetitive DNA; (c) the sequence with masked genes; and (d) the part remaining after removing both repetitive DNA and gene sequences. The last fraction can be referred to as "genome dark matter" (see also Yamada et al. 2003; Bejerano et al. 2004; Ponting and Lunter 2006; Woolfe and Elgar, 2008 for a similar use of this astrophysics terminology in genomics). Obviously, the dark matter term does not imply, in any form, absence of functionally. We compare here the entire human and primate genomes and the aforementioned three types of genome sub-sequences. In addition to the expected good correspondence of human and monkey genomes sequences, we found that the three genome sub-spaces also show good correspondence. Re-enumeration of the macaque chromosomes based on our mdot-plot tracks for the whole genome sequence resulted in a further increase of correspondence of the two genomes that holds also in the three sub-spaces.

## Materials and Methods

### Materials

In this study we considered genome sequences of five species: humans, three apes (chimpanzee, gorilla and orangutan), and macaque. For each genome, four types of sequences were used: (a) the whole genome and the sequences with (b) masked repeated DNA; (c) masked genes; and (d) the part remaining after masking both repeated DNA and genes (the part that we call "genome dark matter"). The chromosomes were divided into 1Mb length segments ordered in the natural way within each chromosome; the chromosomes were ordered, for the sake of certainty, in ascending order with respect to their numbers, chromosomes X and Y being at the end of the list. We consider each segment as a "point", so that the genome is transformed into a sequence of points. Thus, the human genome with about 3,000 million bps is transformed into a sequence of 3,000 points in length.

**Distance between sequences**

Further consideration requires a definition of distance between any pair of segments. An approach for comparing long genome segments, based on the frequency distribution of short oligonucleotides (k-mers) which occur in the segments, was developed quite a long time ago (Brendel et al. 1986; Karlin and Ladunga 1994; Kirzhner et al. 2002; Rocha et al. 1998; Wu et al. 1997). We refer to this technique as *compositional spectra* (*CS*) analysis. This method was employed in a large body of studies on phylogenetic relationships (e.g., Kirzhner et al. 2003, 2005; Sims et al. 2009a). Versions of the method differ mainly in the choice of the set of oligonucleotides, referred to as *support*, for which the frequency distribution (CS) is evaluated. Three versions of the method are employed in the present work, which differ in the support: (i) 10-letter words (10-mers) in the four letter alphabet; (ii) 6-letter words (6-mers) in the four letter alphabet; and (iii) 20-mers in a two letter purine-pyrimidine alphabet. In our setup, 10-mers and 20-mers may occur in a sequence under consideration even in the case of mismatch at two positions ($r=2$), while 6-mers must occur with zero mismatch ($r=0$). In our tests, the support size chosen was equal to 256 words for versions (i) and (ii) and 250 words for version (iii) (see Kirzhner et al. 2002, 2003 for more details). The method of calculating the distances between the CS were chosen with regard to the fact, established in Kirzhner et al. 2002; Volkovich et al. 2005 that the relative distances between different DNA sequences substantially depend on the chosen type of distance function. In particular, it was shown that using Euclidian distance results in less variable pair-wise distances between species compared to the distances based on the Spearman rank correlation coefficient (Kendall 1970). Using rank correlation for the three versions of analysis (i-iii) generates three distances for any pair of compared sequences. The fourth distance employed in the analysis was the absolute difference between the sequences in the GC content. For the sake of statistical stability, we calculate the distance between two segments only if the total length of the masked regions for each segment does not exceed 20%. The number of such segments for the dark matter of the human genome, for example, was about 1800 (the total number of segments was about 3000).

**Dot-plot analysis**

In all four distances, the set of pair-wise distances between segments includes many values (actually from a continuum), preventing direct use of the classical dot-plot analysis for the comparison of genomes. Therefore, the analysis here is based on our modification of the dot-plot method for continual distance values referred to as mdot-plot (Kirzhner et al. 2011). With this approach, locally-minimal elements should first be determined in the distance matrix (see Procedure 1 in Kirzhner et al. 2011). For the set of such elements (T-set) we then define tracks, i.e., sequences of dots parallel to the bisector of the plot. As with the standard dot-plot method, these tracks allow the detection of extended regions of similarity of the compared genomes. In the current analysis, we have four T-sets: $T=T_{CG}$, $T_6$, $T_{10}$, and $T_{RY}$, corresponding to the four employed distances. We will also consider tracks corresponding to more stringent conditions of similarity by using different intersections of the T-sets. Namely, define as a T-set, the set of points consisting of a certain combination of sets $T_6$, $T_{10}$, $T_{RY}$, $T_{CG}$. We designate $\mathbb{T}_4 = T_6 T_{10} T_{RY} T_{CG}$, where the product means the intersection of the sets. Likewise, define $\mathbb{T}_3$ as a set of points which includes only those points that belong simultaneously to at least three out of the four sets: $\mathbb{T}_3 = T_{CG}T_6T_{10} + T_{CG}T_6T_{RY} + T_{CG}T_{10}T_{RY} + T_6T_{10}T_{RY}$. Similarly, $\mathbb{T}_2$ will be the set that includes only points that belong simultaneously to at least two out of the four sets: $\mathbb{T}_2 = T_{CG}T_6 + T_{CG}T_{10} + T_{CG}T_{RY} + T_6T_{10} + T_6T_{RY} + T_{10}T_{RY}$. We also consider the set $\mathbb{T}_1 = T_{CG} + T_6 + T_{10} + T_{RY}$. It is clear that genuinely similar segments should be close by all distances, while detected random similarities will be filtered out by the foregoing compound criteria.

## Results and Discussion

**1. The main pattern**

In Fig.1 we compare the human genome with three ape genomes and the macaque genome based on the mdot-plot diagram. One can easily see a considerable concentration of tracks (of length ≥ 4 points) along the main bisector, which implies a similar organization of the compared genomes. We will call this abundance of tacks in the bisector region as "the main pattern". Keeping in mind that the numeration of ape chromosomes relative to human chromosomes corroborates with the conserved order of orthologous genes (parallel synteny), we can conclude that our mdot-plot analysis of the above-gene-level sequence organization remarkably reveals the same ordering. Moreover, we found the same ordering when genes were masked in the targeted sequences (Fig. 1). Unlike human-apes diagrams, the human-macaque comparison produced a diagram with tracks scattered over the plane, despite the relatively large track sizes. A more detailed mdot-plot comparison of macaque and human genomes is presented in Fig. 2. Here the whole sequence was partitioned into chromosomes. Although the tracks are scattered, quite often they are located within the human-macaque chromosomal boxes. By rearranging of whole macaque chromosomes we were able to get the tracks near (or in) the bisector. Thus, tracks associated with macaque chromosomes 12 and 13 appeared to fully correspond to human chromosome 2 (see Fig. 2a). Transposing macaque chromosomes 12 and 13 instead of its chromosome 2, leads to a bisector location of their tracks. Further transpositions result in a new mdot-plot with the tracks concentrated near the bisector (Fig. 2b). Corresponding re-numeration of macaque chromosomes (Fig. 2c) associated with this analysis nearly perfectly coincides with the order based on the analysis of parallel synteny of conserved microsatellite loci (Rogers et al. 2006). We employ this order in the further considerations of macaque-human genome correspondences.

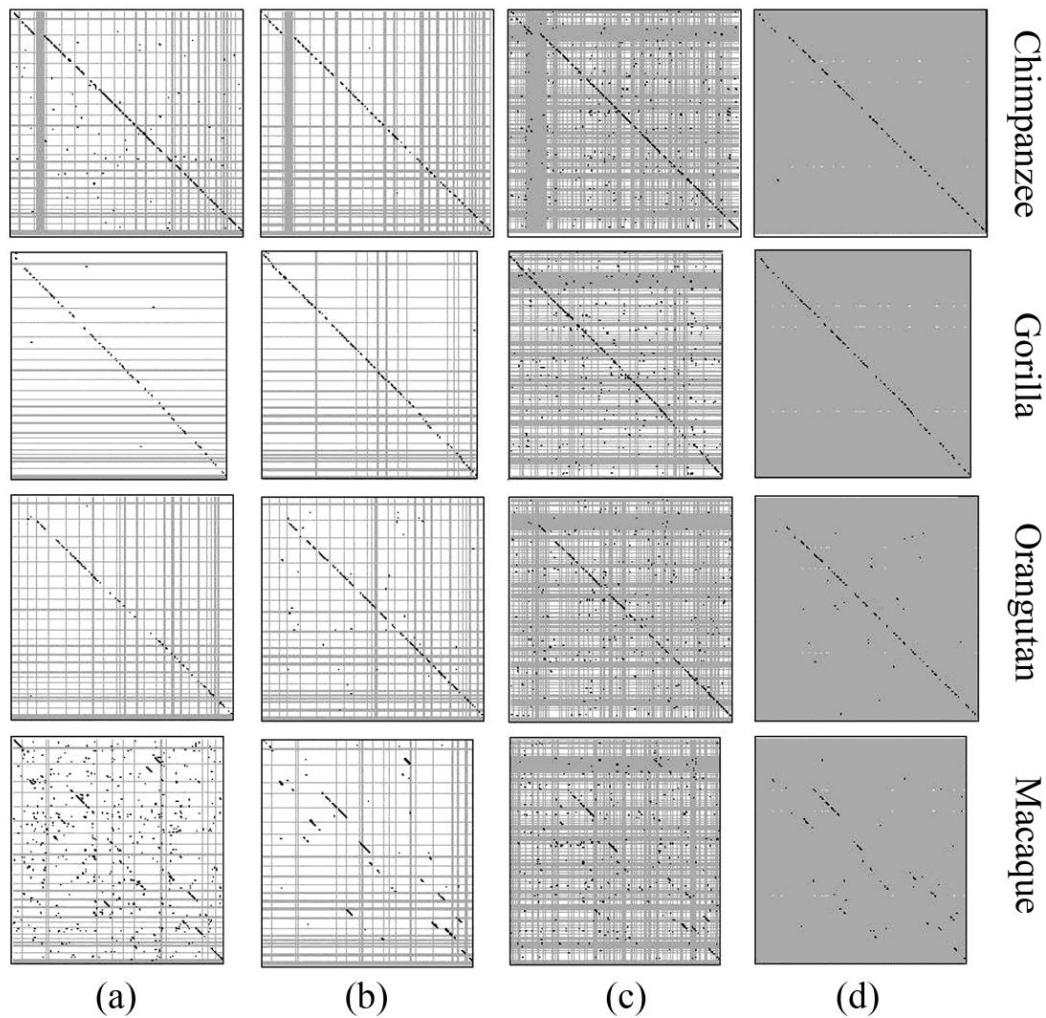

**Fig. 1. Mdot-plot of human, apes and macaque genomes for different fractions of genome sequences.** In all cases, mdot-plot was built on the set of points $T_4$. Axes X and Y represent coordinates in monkey and human genomes, respectively. (a) the whole genome sequence; (b) masked repetitive sequences; (c) masked gene sequences; (d) dark genome matter. Shown are tracks of lengths ≥4; grey stripes correspond to coordinates with either missing or masked sequences.

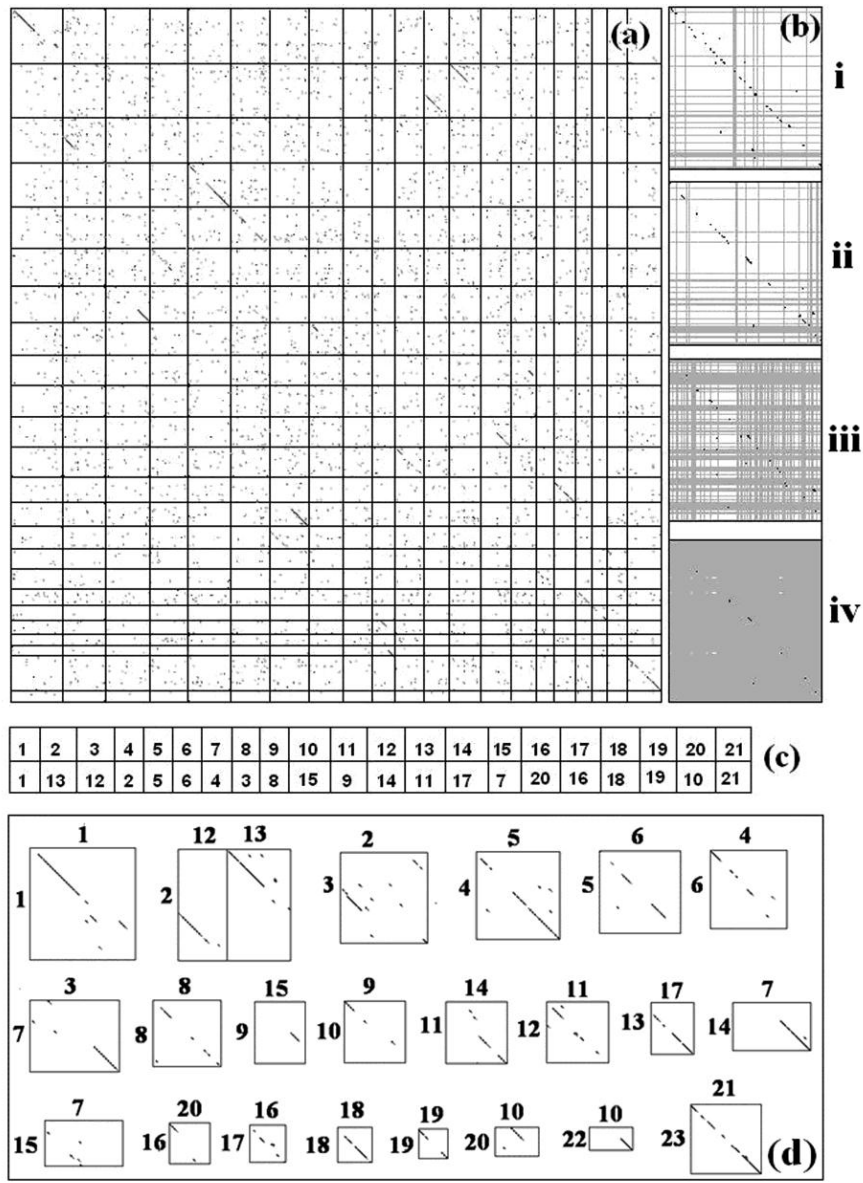

**Fig. 2. Macaque-human mdot-plot diagram for track length size ≥4**: (a) mdot-plot with standard enumeration of macaque chromosomes for $T_{CG}$ set (X- chromosome is the last in the list, numbered as 21, Y-chromosome was not included); (b) mdot-plots with $T_4$ set: (i) for full genome sequence, (ii) for genome with masked repeats, (iii) for genome with masked genes, (iv) for genome dark matter, with the new ordering of macaque chromosomes; (c) new ordering of macaque chromosomes corresponding to the main pattern b; (d) whole-genome mdot-plot analysis based on $T_3$ set.

The order of macaque chromosomes presented in Fig. 2c allows for obtaining the main pattern in the mdot-plot diagram, when the similar (presumably, orthologous) regions of human and macaque genomes are concentrated in the vicinity of the bisector. A more detailed comparison of human and macaque genomes corresponding to Fig. 2a is shown in Fig. 2d. Here the mdot-plot analysis was based on $\mathbb{T}_3$ set. Considerable segments of parallel synteny can be seen in many cases while for some genomic regions (e.g., human chromosome 15 and macaque chromosome 7) it is rather fragmentary. The correspondence is expressed more clearly if we relax the conditions of coincidence of minimal points (e.g., by using $\mathbb{T}_2$ or even $\mathbb{T}_1$).

## 2. Reliability of the main pattern

Our algorithm for track detection is based on a set T of local minima of the distance matrix. Two mechanisms were employed to reduce the effect of chance on track appearance: (a) using all of the four distances in searching for local minima, and (b) filtering out too short tracks. Obviously, the probability of obtaining a long track by chance is very low, and even more so is the chance of obtaining dozens of long tracks. Consider the possibility of using simultaneously the two filtering mechanisms (a & b) in the analysis of similarity of genomic sequences, based on the example of comparison of complete human and chimpanzee genomes (Fig. 3). .A detailed examination of the ratio of tracks inside and outside of the selected zones demonstrates that with the transition from $T_{CG}$ to a more stringent rule based on $\mathbb{T}_4$, the set of tracks of considerable length (>5) remains virtually unchanged, unlike the tracks outside the described zones, that tend to vanish. Thus, even one T set allows to reveal the main pattern, but, as expected, the signal/noise ratio improves with the number of simultaneously employed T sets, i.e., with the stringency of the selection rule.

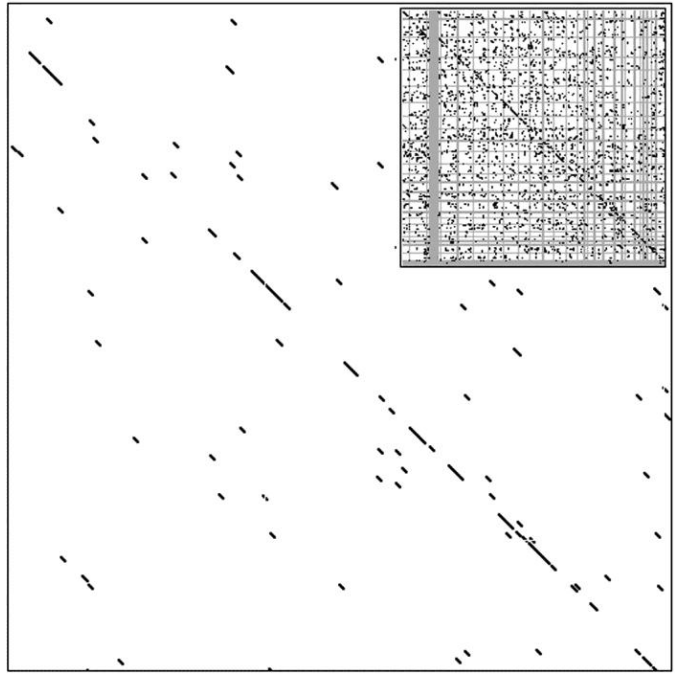

**Fig. 3 A fragment of mdot-plot diagram of full genome sequences of human and chimpanzee built for GC distance.** Shown are tracks with lengths ≥4; the box in the upper-right corner represents the full diagram.

Let us consider the length distribution of the tracks and their positioning relative to the bisector. We start with the least stringent selection rule in the mdot-plot procedure by using the $\mathbb{T}_1$ set (Fig. 4A). While tracks with length up to 5-6 points can be found at any distance from the bisector, longer tracks are concentrated in relatively narrow zones. Thus, in the human-pan histogram we find two such zones: at distances of 0-50 and 100-150 points from the bisector (Fig. 4A-i). For human-gorilla and human-macaque comparisons, long tracks are found in the zone 0-100 around the bisector (Fig. 4A-ii,iii) and in the 0-50 zone for human-orangutan (Fig. 4A-iv). When increasing the stringency of the selection rule in the mdot-plot analysis, one expects to reduce the proportion of noisy (random) tracks together with preservation of the long tracks near the bisector. This is exactly what we observed in the human-chimpanzee mdot-plot using rather liberal T=$T_{CG}$ and much more stringent $\mathbb{T}_4$ (which also includes $T_{CG}$) (Fig. 4B). A comparison of the two parts of the figure (i and ii) shows a clear reduction in the abundance of short tracks when the $T_4$ set is applied. Thus, tracks with $l=4$ appear inside the bisector zone, 199 and 42 times, for T=$T_{CG}$ and T=$\mathbb{T}_4$, respectively (a 5-fold reduction). For the outside zone, corresponding numbers will be 1984 and 7 (more than 280-fold reduction). Another finding is that longer tracks detected by $T_{CG}$ inside the bisector zone tend also to survive under $\mathbb{T}_4$. In general, assuming a "random model" for appearance of long tracks, one should expect to observe a high frequency of their destruction during transition from the $T_{CG}$ to $\mathbb{T}_4$ detection rule, but this was not the case. Therefore, we consider the resistant mdot-plot tracks with a length of 10-25 points (i.e., 10-25 Mpb length) as objective indicators of local genome similarity beyond parallel syntenic chains of structural genes.

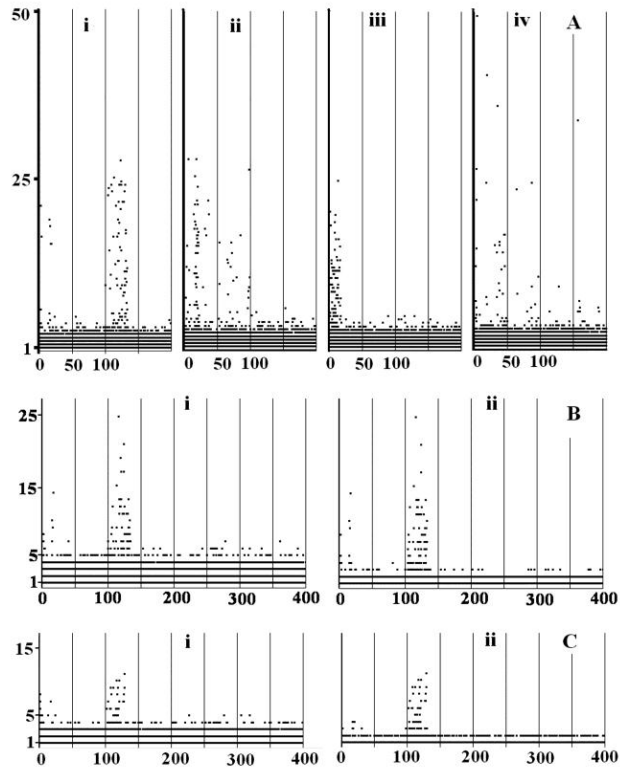

**Fig. 4. Length distribution of the tracks and their positioning relative to the bisector in the human-primates mdot-plot comparison.** Tracks of length $l$ (Y axis) located at distance $x$ from bisector (X axis) are denoted by dots (thus the multiplicity is not taken into account). **(A)** Set $T_1$ was used for full genome mdot-plot analysis: (i) human-chimpanzee; (ii) human-gorilla; (iii) human-orangutan; (iv) human-macaque. **(B)** human-chimpanzee full genome comparison: (i) using $T_{CG}$; (ii) using $T_4$. **(C)** human-chimpanzee of genome dark matter comparison: (i) using $T_{CG}$; (ii) using $T_4$.

An interesting question is whether the results of comparisons of the whole genome sequences are also valid the genome dark matter (the part remaining after masking coding DNA and repeated elements). The resulting histogram of length distribution of the tracks and their positioning relative to the bisector (Fig. 4C) proved very similar to the histogram in Fig.4B for the whole genome sequence comparison. We find that the longest tracks appear in the same two zones around the main bisector as in the case of whole genomes, although the tracks are shorter compared to those for the whole genomes. One possible explanation of this result is the absence of sufficiently large segments after sequence masking. Alternatively, one should take into account that the compositional spectra of full sequences and sequences remaining after masking may differ, hence the differences in track length distribution. The results in Fig. 4C demonstrate that with the transition from $T_{CG}$ to $\mathbb{T}_4$, the set of tracks of considerable length (>6) remains virtually unchanged, unlike the tracks outside the described zones.

As with the whole genome sequence comparisons, the analysis of dark genome matter shows that with increasing stringency of the selection rule, the track lengths in the outside region decrease much faster than in the inside regions. However, unlike the whole-genome comparisons, the track lengths do not increase with the selection stringency. This difference may result from two reasons: (1) Masking in general can change the compositional spectra of the compared segments; the correspondingly changed distance matrix may not necessarily coincide with the properties of the distance matrix obtained for unmasked segments (e.g., due to disappearance of long tracks); or (2) masking does not leave sufficiently long continuous sections in the genomic sequence. Recall that we calculate the distance between two segments only if the total length of the masked regions for each fragment does not exceed 20%; otherwise the fragment is considered as "completely masked" (gap). By definition, a track cannot include a completely masked fragment; hence the track lengths depend on the distribution of such gaps in the considered sequence of 3000 (the whole genome).

The last reason does indeed affect the track lengths. To demonstrate that, we calculated the set of tracks for the full genomic sequence using T= $\mathbb{T}_4$ and then masked the set of tracks in accordance with the masking conditions applied for defining dark matter sequence. The results of the comparison of the masked set of tracks with initial dark matter tracks are presented in Table 1. Although the maximal tack length for unmasked genomes was 25 (see Fig. 4B), after the conducted track masking the maximum length of track is 11 as it was in the case of initial mdot-plot analysis of dark matter sequences. This means that the long tracks observed in the full genome were then "destroyed" by the track masking procedure.

**Table 1.** Comparing two variants of dark matter track analysis using $\mathbb{T}_4$ set

| Length of tracks | 3 | 4 | 5 | 6 | 7 | 8 | 9 | 10 | 11 |
|---|---|---|---|---|---|---|---|---|---|
| (A) Track analysis of the full sequence followed by masking | | | | | | | | | |
| Inside | 37 | 16 | 12 | 3 | 4 | 5 | 2 | 2 | 1 |
| Outside | 10 | 0 | 0 | 0 | 0 | 0 | 0 | 0 | 0 |
| (B) Masking of the full sequence followed by track analysis | | | | | | | | | |
| Inside | 43 | 16 | 3 | 8 | 4 | 3 | 4 | 2 | 1 |
| Outside | 13 | 0 | 0 | 0 | 0 | 0 | 0 | 0 | 0 |

## 3. GC distribution and fragment similarity

If our goal is to find segments in human and ape genomes that have not changed much in the short evolutionary time since their common ancestor, would it be not enough to confine ourselves to comparing the similarity of CG only, because close sites must have similar GC? As we could see from the results displayed in Fig.3, 4B-i and Fig.5 comparing full genomes by using GC-based distance indeed gives reasonable results. However, combining GC-content and 20-mer words in RY alphabet dramatically improves detectability of the main pattern (Table 2). This means that important information on similarity of compared sequences is present in the abundance of oligonucleotide words even if a partial reduction of the information was caused by use of the "degenerative" alphabet. We employed here RY alphabet in order to make sure that the effect of the word construction is not directly influenced by GC content. Still, combining CG-content and 6- and 10-mers words in A-T-G-C alphabet gives a similar improvement of detectability, confirming the importance of word abundances (which reflect letter order in addition to letter proportions) in detecting similarly organized segments in the compared genomes (see Table 2). An additional illustration of the role of letter order in the detection of segment similarity is provided by the results of the mdot-plot analysis conducted separately with each of the three vocabularies and their combination (Fig. 5) without including GC-content to the selection rule. In each case, the detectability of long tracks in the bisector region is increased compared to the case of using the pure GC-based rule. In general, this result is quite expected. Obviously, two genomic segments with very similar GC content may strongly differ with respect to sequence organization. Correspondingly, GC- and CS-distances may be poorly correlated. Indeed, using bacterial genomes we showed earlier that random permutations of G and C (or A and T) letters without changing S (G or C) and W (A or T) positions result in significant changes of CS-distances (Kirzhner et al. 2007). The results presented in Table 2 can be considered as another demonstration of complementary contribution of positional variation in GC content and k-mer abundances to (evolutionary) distances of the compared sequences.

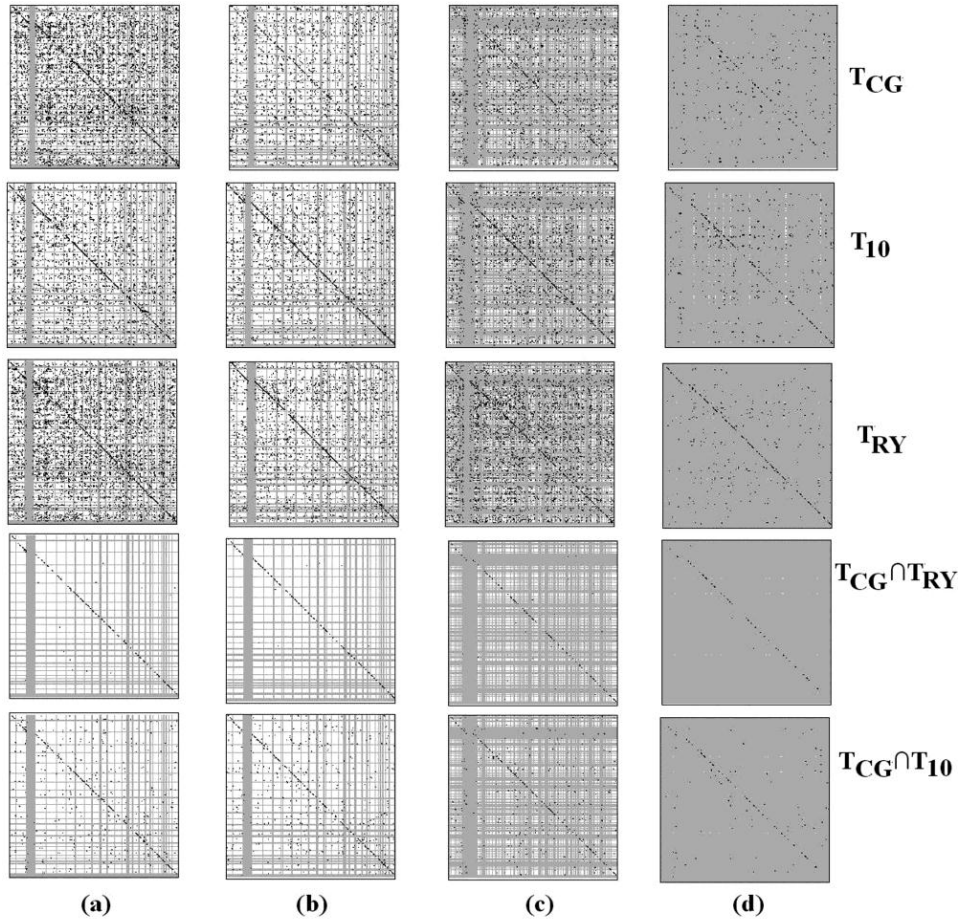

**Fig. 5. Mdot-plots for different fractions of human and chimp genome sequences constructed using different T-sets**: T=$T_{CG}$, $T_{10}$, $T_{RY}$, $T_{CG} \cap T_{RY}$, and $T_{CG} \cap T_{10}$ (compare with the results in Table 2). (a) the whole genome sequence; (b) masked repetitive sequences; (c) masked gene sequences; (d) dark genome matter. Tracks with $l_{min} \geq 4$ are shown.

An interesting question is how this complementarily is manifested on different fractions of the genome sequence. As one would expect, after masking repetitive elements, the detectability of our general pattern (long tracks in the region spanning the bisector) increases, compared to the analysis of the full genome, for all considered T sets (Fig. 5a,b). A more important fact is that the general pattern remains after masking the gene sequences (Fig. 5c) or both gene sequences and repetitive DNA implying considerable phylogenetic information content of genome dark matter (Fig. 5d). It is noteworthy that the ranking of different distances with respect to their efficiency in revealing the main pattern remains qualitatively similar on the different genome fractions (compare the ranking of the columns a-d on each of the rows in Fig. 5).

**Table 2.** Effect of distances combination in the detection of segments similarity Human-Pan full genomes.

| Track length, $l$ | $T=T_{CG}$ | | $T=T_{CG} \cap T_{RY}$ | | $T=T_{CG} \cap T_6 \cap T_{10}$ | | $T=\mathbb{T}_4$ | |
|---|---|---|---|---|---|---|---|---|
| | Inside | Outside | Inside | Outside | Inside | Outside | Inside | Outside |
| 3 | 973 | 12110 | 57 | 311 | 183 | 1625 | 53 | 153 |
| 4 | 199 | 1984 | 46 | 16 | 58 | 158 | 42 | 7 |
| 5 | 43 | 331 | 11 | 1 | 14 | 13 | 10 | 0 |
| 6 | 13 | 58 | 7 | 0 | 8 | 3 | 7 | 0 |
| 7 | 15 | 10 | 10 | 0 | 13 | 0 | 10 | 0 |
| 8 | 9 | 1 | 7 | 0 | 9 | 0 | 7 | 0 |
| 9 | 7 | 2 | 7 | 0 | 7 | 0 | 7 | 0 |
| 10 | 2 | 1 | 2 | 0 | 2 | 0 | 2 | 0 |
| 11 | 6 | 0 | 7 | 0 | 8 | 0 | 7 | 0 |
| 12 | 4 | 0 | 4 | 0 | 4 | 0 | 4 | 0 |
| 13 | 4 | 0 | 3 | 0 | 4 | 0 | 3 | 0 |
| 14 | 1 | 0 | 1 | 0 | 1 | 0 | 1 | 0 |
| 15 | 1 | 0 | 1 | 0 | 1 | 0 | 1 | 0 |
| 16 | 0 | 0 | 0 | 0 | 0 | 0 | 0 | 0 |
| 17 | 2 | 0 | 1 | 0 | 1 | 0 | 1 | 0 |
| 18 | 0 | 0 | 0 | 0 | 0 | 0 | 0 | 0 |
| 19 | 1 | 0 | 0 | 0 | 0 | 0 | 0 | 0 |
| 20 | 0 | 0 | 0 | 0 | 0 | 0 | 0 | 0 |
| 21 | 1 | 0 | 1 | 0 | 1 | 0 | 1 | 0 |
| 22 | 0 | 0 | 0 | 0 | 0 | 0 | 0 | 0 |
| 23 | 0 | 0 | 0 | 0 | 0 | 0 | 0 | 0 |
| 24 | 0 | 0 | 0 | 0 | 0 | 0 | 0 | 0 |
| 25 | 1 | 0 | 1 | 0 | 1 | 0 | 1 | 0 |

Although repetitive elements can be considered as a noisy part of the genome in human-monkey comparisons, our mdot-plot analysis based on compositional spectra was able to reveal a certain manifestation of the main pattern when only repetitive sequences remained in the compared human-chimpanzee genomes (Fig. 6a). The usual practice of Blastz analysis of whole genome comparisons is to mask repetitive elements as a source of noise (Schwartz et al. 2003); this noise effect is also expressed in our analysis (compare columns a and b in Fig. 5). Keeping

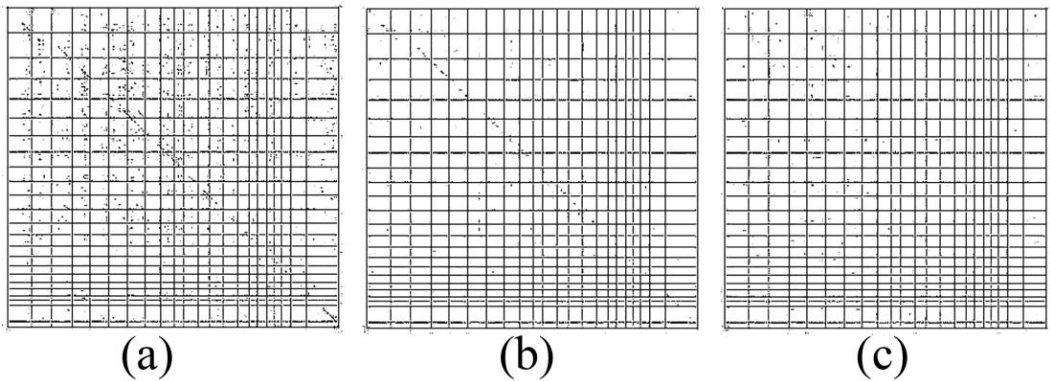

**Fig. 6 Mdot-plot diagrams for comparison of genome subsequences** (Tracks with $l_{min}$ ≥4 are shown): (a) human-chimpanzee genic space; (b) human-chimpanzee repetitive sequences; (c) human genic space vs. human repetitive sequences. The distances were calculated based on CS-analysis of 10-mers

this in mind, a remarkable fact is that taken along as a separate genome fraction, repetitive sequences displayed an obvious signal of conservation between human and chimpanzee, despite a considerable noisy component (compare the mdot-plots for genic space and repetitive sequences in Fig. 6, a and b). It is noteworthy that the revealed similarities cannot be considered as a result of a general "genome dialect" (Paz et al. 2006) shared between human and pan. Indeed, even if the considered fractions (genic sequences and repetitive sequences) are taken from the same genome, they do not display chromosome ordering pattern (Fig. 6c). In general, we can conclude that fuzzy similarity (parallel synteny) between human and chimpanzee genomes extends beyond chains of highly conserved noncoding elements and orthologous genes and is detectable by CS-based mdot-plot analysis on genome dark matter and repetitive sequences. The importance of such fuzzy similarities was also hypothesized by other authors (McLean and Bejerano 2008; Sims et al. 2009b).

## 4. Density of genome coverage

As expected, with increasing stringency of the selection rule, the track lengths in the outside region decrease much faster than in the inside region. For each of the sets $\mathbb{T}_1$- $\mathbb{T}_4$ we can separately define a minimal length of tracks ($l_{min}$) that appear in the inside region corresponding to the set T while all outside tracks are shorter than this minimum. By taking out tracks of smaller size, the mdot-plot diagrams presented in Fig. 7 (part A) will be obtained. These results show that by combining T sets with consequently decreasing stringency and corresponding minimum track lengths with increasing $l_{min}$ (i.e., decreasing stringency) we can obtain distinct and apparently reliable patterns of similarity of genomes. The mdot-plot diagrams in Fig. 7 (part A) can also be represented in another form: as a projection reflecting the "coverage" of human genome by parallel syntenic regions of the pan genome (Fig. 7, part B). Comparing of parts i and ii parts in Fig. 7 allows concluding that decreasing stringency of T sets coordinated with selection of longer tracks leads to higher coverage and, simultaneously, to a lower proportion of undesirable regions with double coverage. As expected, the most liberal selection rule gives the highest coverage (part iv in Fig. 7). Using this feature as a selection rule, we extended the analysis to other human-monkey comparisons and different genome fractions (Table 3). As expected, for all comparisons, the best coverage was found after masking repetitive elements (i.e., when the combined fraction "genes+dark matter" was targeted). Remarkably, even without the genic fraction, the dark matter sequences give a rather good coverage, implying a considerable "fuzzy" conservation that extends far beyond genic space (Dermitzakis et al. 2005; Khalil et al. 2009). This corroborates well with our earlier conclusions and the findings of other authors that alignment-free analysis is able to reveal reliable phylogenetic information in whole-genome sequence comparisons (Blaisdel 1989; Karlin et al. 1998; Kirzhner et al. 2002, 2007; McLean and Bejerano 2008; Sims et al. 2009a). It is also worth mentioning that very similar coverage was obtained in human-monkey comparisons with Blastz-net/Lastz-net software (http://useast.ensembl.org/info/docs/compara/analyses.html): 0.73 for Human-Chimpanzee and 0.64-0.65 for the remaining three considered pairs.

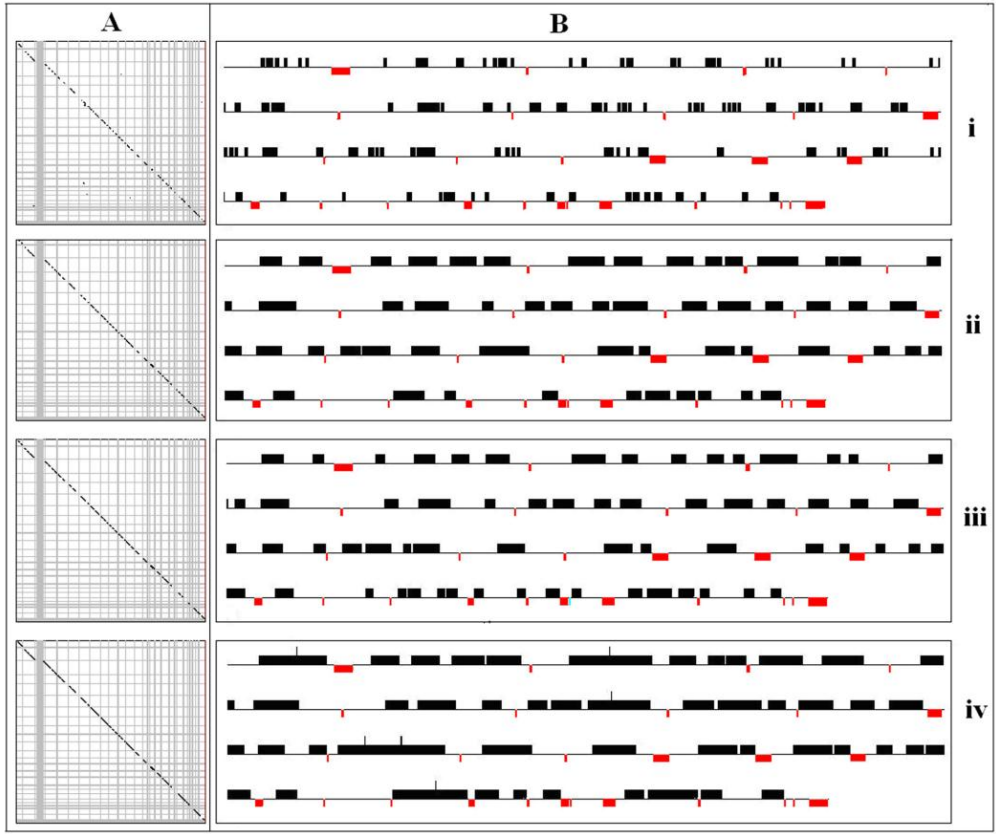

**Fig. 7. Mdot-plot analysis for human-chimpanzee genome comparisons.** (a) set $T_4$ with $l_{min} \geq 4$; (b) set $T_3$ with $l_{min} \geq 9$; (c) set $T_2$ with $l_{min} \geq 13$; (d) set $T_1$ with $l_{min} \geq 14$. Left: mdot-plot diagrams; right: the scheme of human genome coverage by segments of chimpanzee genome (covered regions are marked by black bars whereas double-coverage regions are marked by additional thin upper bars, corresponding to the size of regions; the bottom bars correspond to "no data" segments in the human genome sequence).

## Conclusions and prospects

Here we have shown the efficiency of compositional spectra analysis based on k-mer scoring in the DNA sequences as an efficient tool for comparative genomics. We employed the distance matrix that includes pair-wise distances of each-versus-each segments of the compared genomes. Therefore, the revealed conserved syntenies are based on fuzzy similarity rather than on chains of discrete anchors (genes or highly conserved noncoding elements). The revealed similarity segments extend from 4 to 15 and more Mbp. In addition to the good correspondence found in human-apes comparisons of whole-genome sequences, which was expected because of close taxonomic relations, a good similarity was also found after masking gene sequences,

**Table 3.** Mdot-plot analysis of human genome coverage by different fraction of monkey genome

|  | (a) | | | (b) | | | (c) | | | (d) | | |
|---|---|---|---|---|---|---|---|---|---|---|---|---|
|  | 1 | 2 | 3 | 1 | 2 | 3 | 1 | 2 | 3 | 1 | 2 | 3 |
| Pan ($\mathbb{T}_4$) | 3 | **0.30** | 0.03 | 3 | **0.33** | 0.02 | 3 | **0.23** | 0.02 | 3 | **0.25** | 0 |
| Pan ($\mathbb{T}_1$) | 10 | **0.69** | 0.02 | 9 | **0.75** | 0.03 | 9 | **0.55** | 0.02 | 6 | **0.54** | 0.03 |
| Gorilla ($\mathbb{T}_4$) | 3 | **0.22** | 0.02 | 3 | **0.32** | 0.01 | 3 | **0.15** | 0.03 | 3 | **0.18** | 0 |
| Gorilla ($\mathbb{T}_1$) | 10 | **0.72** | 0.02 | 10 | **0.79** | 0.02 | 9 | **0.52** | 0.02 | 7 | **0.54** | 0 |
| Pongo ($\mathbb{T}_4$) | 3 | **0.22** | 0.04 | 3 | **0.22** | 0.05 | 3 | **0.11** | 0.05 | 3 | **0.18** | 0.05 |
| Pongo ($\mathbb{T}_1$) | 11 | **0.52** | 0.07 | 10 | **0.54** | 0.06 | 9 | **0.33** | 0.07 | 8 | **0.32.** | 0.04 |
| Macaque ($\mathbb{T}_4$) | 3 | **0.17** | 0.02 | 3 | **0.16** | 0. | 3 | **0,11** | 0.02 | 3 | **0.10** | 0.01 |
| Macaque ($\mathbb{T}_1$) | 9 | **0.41** | 0.01 | 10 | **0.41** | 0.02 | 10 | **0.22** | 0.03 | 8 | **0.41** | 0.03 |

Genome fractions: (a) the whole genome sequence; (b) masked repetitive sequences; (c) masked gene sequences; (d) dark genome matter.

Columns: 1 – minimal length tracks counted for estimating the coverage; 2 – coverage fraction; 3 – double-coverage fraction.

indicating that the CS-analysis manages to reveal phylogenetic signal in the remaining part of the genome sequences, i.e., in the organization of genome dark matter and repetitive DNA. A similar although a bit more complicated situation was found in human-macaque genome comparisons. Despite good local correspondence between these two species on a chromosomal level, we had to renumber macaque chromosomes to get a nearly full correspondence. Remarkably, our enumeration proved to correspond well with the results on parallel synteny based on mapping microsatellite loci.

   Obviously, the possibility to reveal parallel ordering of the compared sequences derives from the signal in the sequence organization that varies locally along the corresponding segments of the compared genomes and is detectable by the scoring method, e.g., mdot-plot based on k-mer abundances. Here we explored here two sources contributing to this signal: sequence composition (using GC content of the scored segments) and sequence organization (analyzing k-mer abundances in A-T-G-C or R-Y alphabets). As we could see, comparison of full genomes based on GC distribution along the sequences indeed gives reasonable results. However, combining GC-content with k-mers dramatically improves the ordering quality. Therefore, GC-content and k-mer abundances can be considered as complementary sources of information on evolutionary conserved similarity of genome sequences (see also Kirzhner et al. 2007; Mrázek 2009).

An important perspective of the proposed approach is the comparison of distant species where the possibility to detect highly conserved anchor sequences is problematic. For example, the described approach allowed detecting numerous segments of diffused similarity between human and zebrafish genomes of up to seven million base pairs and more. Using such analysis we can search for remnant signals of retained large segmental duplications as indicators of the paleopolyploid nature of extant vertebrate genomes (the problem of "genome halving" - Yin and Harteming 2005; Zheng et al. 2008). The existence of large-scale intragenomic similarities, relevant for most groups of eukaryotes (Dehal and Boore 2005; Freeling and Thomas 2006), so far could be tested based mainly on conserved orders of syntenic orthologous genes and/or chains of ultraconserved noncoding elements,. Extending this formulation to noncoding part of the genome based on the proposed fuzzy approach of large-scale mdot-plot comparisons may be especially useful, keeping in mind the considerable uncertainty in score-based alignments even if probabilistic models are employed for alignment (Lunter et al. 2008).